%% file: cenx4MNRAS.tex
\documentclass[a4paper,fleqn,usenatbib]{mnras}

\usepackage{newtxtext,newtxmath}

\usepackage[T1]{fontenc}
\usepackage{ae,aecompl}

\usepackage{graphicx}
\usepackage{amsmath}
\usepackage{amssymb}
\usepackage{multirow}

\title[The inclination of Cen X-4]{Constraining the inclination of the Low-Mass X-ray Binary Cen X-4}

\author[Hammerstein et al.]{
Erica K. Hammerstein$^{1, 2}$\thanks{ekhammer@umich.edu},
Edward M. Cackett$^2$,
Mark T. Reynolds$^1$,
Jon M. Miller$^1$
\\$^1$ Department of Astronomy, University of Michigan, 1085 South University Ave, Ann Arbor, MI 48109-1107, USA
\\$^2$ Department of Physics \& Astronomy, Wayne State University, 666 W. Hancock St., Detroit, MI 48201, USA
}

\date{Accepted XXX. Received YYY; in original form ZZZ}

\pubyear{2018}

\begin{document}
\label{firstpage}
\pagerange{\pageref{firstpage}--\pageref{lastpage}}
\maketitle

\begin{abstract}
We present the results of ellipsoidal light curve modeling of the low mass X-ray binary Cen~X-4 in order to constrain the inclination of the system and mass of the neutron star. Near-IR photometric monitoring was performed in May 2008 over a period of three nights at Magellan using PANIC. We obtain $J$, $H$ and $K$ lightcurves of Cen~X-4 using differential photometry. An ellipsoidal modeling code was used to fit the phase folded light curves. The lightcurve fit which makes the least assumptions about the properties of the binary system yields an inclination of $34.9^{+4.9}_{-3.6}$ degrees (1$\sigma$), which is consistent with previous determinations of the system's inclination but with improved statistical uncertainties. When combined with the mass function and mass ratio, this inclination yields a neutron star mass of $1.51^{+0.40}_{-0.55}$~$M_{\sun}$. This model allows accretion disk parameters to be free in the fitting process. Fits that do not allow for an accretion disk component in the near-IR flux gives a systematically lower inclination between approximately 33 and 34 degrees, leading to a higher mass neutron star between approximately $1.7~M_{\sun}$ and $1.8~M_{\sun}$. We discuss the implications of other assumptions made during the modeling process as well as numerous free parameters and their effects on the resulting inclination.
\end{abstract}




\section{Introduction}

Low mass X-ray binaries (LMXBs) are a type of binary system in which a compact object, a neutron star or black hole, is accompanied by a secondary star of lower mass.  Since the compact object is in a binary system its mass can be measured.  Neutron star masses are of particular interest since good constraints on neutron star masses and radii in turn put constraints on the equation of state for dense matter \citep[e.g.][]{steiner13}.  Massive neutron stars can strongly constrain the equation of state \citep{lattimer11}, with the most massive known to date having a mass of $2.01\pm0.04$~M$_\odot$ \citep{antoniadis13}.  Neutron stars in low-mass X-ray binaries can gain mass through accretion, and hence may be expected to be more massive than the canonical 1.4~M$_\odot$.

Orbital variability of LMXBs is used to determine the binary masses. The secondary star fills its Roche lobe, leading to mass transfer to the compact object, but also meaning that the secondary has an ellipsoidal, tear-drop, shape due to the gravitational influence of the compact object.  The 
brightness of the secondary star therefore changes throughout the binary orbit, and modeling of these ellipsoidal variations leads to a measure of the binary inclination, $i$.
Moreover, the radial velocity curve of the binary system can be used to determine the mass function, $f(M)$:
\begin{equation}
	f(M) = \frac{P_{\rm orb}K^3_2}{2\pi G} = \frac{M_1sin^3_i}{(1+q)^2}
	\label{eq:fm}
\end{equation}
where $P_{\rm orb}$ is the orbital period, $K_2$ is the radial velocity semi-amplitude, $M_1$ the mass of compact object, and $q = M_2/M_1$ the mass ratio of the secondary star and compact object. Combining the inclination with the mass function yields the mass of the compact object. In this paper, we model the ellipsoidal variations to improve the measured inclination and therefore the mass of the neutron star in the LMXB Cen~X-4.

The compact object in Cen X-4 is a neutron star, due to the observation and identification of a Type I X-ray burst by \citet{matsuoka80}. It is one of the nearest LMXBs \citep[$1.2\pm0.3$~kpc,][]{chevalier89}, allowing for accurate photometric measurements and spectral analyses.  Since its discovery by the Vela~5B satellite in 1969 \citep{conners69}, Cen X-4 has been well studied and many orbital parameters of the system have been determined. The secondary star is estimated to be a K4 V type star with $T_{e\mathit{ff}}$ = 4500 K and $\log g = 3.9$ (\citealt{gonzalez05}, though see \citealt{khargharia10}). \citet{shahbaz14} determined $K_2$ = 147.3 $\pm$ 0.3 km s$^{-1}$, $P_{\rm orb} = 0.629059 \pm 0.000017$~d, and $T_0 = {\rm HJD}~2454626.6214 \pm 0.0002$, 
which are used in the calculations and modeling in this paper, along 
with the mass function, $f(M)$, equal to $0.201 \pm 0.004$~M$_{\odot}$ obtained by \citet{davanzo05}. \citet{shahbaz14} also determined the mass ratio $q = 0.1755 \pm 0.0025$ and inclination $i = 32\degr\substack{+8\degr \\ -2\degr}$.

Previous studies of the system have used several different methods to 
determine the system's inclination and mass ratio.
The first attempt at constraining the system's inclination angle was 
made by \citet{shahbaz93} who fit the quiescent light curve with an
ellipsoidal model, which assumed no contamination from the accretion disk, 
and was determined to be in the range of 31\degr--54\degr. \citet{khargharia10} revised this estimate by determining the contribution of the secondary star in the infrared to remodel the 
original \citet{shahbaz93} light curves, and determined $i$ to be $35\degr\substack{+4\degr \\ -1\degr}$.
\citet{casares07} compared the observed spectrum of the secondary star with a template star
combined with a limb-darkened rotation profile to find $q$. Most recently, \citet{shahbaz14}
determined $i$ and $q$ by using an X-ray binary model to fit the shape of the secondary star's
Roche-lobe distorted absorption line profiles.  The advantage of this approach is that, unlike modeling of ellipsoidal lightcurves, it does not require assumptions about contamination from the accretion disk, and thus provides an independent measure of the inclination.

In this paper, we present near-IR photometry of Cen~X-4 in order to provide an improved measure of the inclination by modeling the ellipsoidal variations, and thus measure the mass of the neutron star.  We present the observations in section 2, and describe the lightcurve modeling in section 3. In section, 4 we discuss the results and systematic effects on the measured inclination based assumptions made in the modeling.  Finally, we present our conclusions in section 5.

\section{Observations \& Data Reduction}

\subsection{Observations with Magellan}

We obtained photometry of Cen~X-4 using the Walter Baade 6.5-m Magellan telescope at Las Campanas Observatory, over three nights: 2008 May 23 -- 25 UT (MJD 54609 -- 54611). The Persson's Auxiliary Nasmyth Infrared Camera (PANIC) was used to observe Cen X-4 in the $J$, $H$, and $K$ bandpass filters, covering 1 -- 2.5 microns \citep{martini04}. PANIC has a pixel scale of 0.125\arcsec, and a $2\arcmin \times 2\arcmin$ field of view. The infrared is best for observing the ellipsoidal variability of X-ray binaries as there is expected to be less contamination from the accretion disk at these wavelengths as opposed to visual wavelengths \citep[up to $\sim$10\% in the $K$-band compared to 50\% in the optical, e.g.,][]{chevalier89, shahbaz93, khargharia10}. The typical exposure time with PANIC was 10 seconds taken in a dice-5 pattern.  Most exposures were taken in $K$, since it has the longest wavelength and should have the least accretion disk contamination; however, a smaller number of exposures in $J$ and $H$ were also taken in order to have some color information.  We obtained 113 useful exposures in $K$, and 19 useful exposures in both $J$ and $H$. A series of darks and twilight flats were taken at the beginning and end of each night.  Observing conditions were excellent over the 3 nights, with typical seeing of 0.5\arcsec.

\subsection{Data Reduction \& Photometry}

Data reduction and analysis used a combination of both the PANIC-specific python scripts\footnote{http://code.obs.carnegiescience.edu/panic} and the astropy python package \citep{astropy}.  We performed calibrations of the images taken, which included dark frame corrections and flat-fielding using twilight flats.  Bias subtraction is unnecessary with PANIC.  Each series of 5 images taken in the dice-5 pattern were sky-subtracted and added together before performing photometry.

\begin{figure}
\centering
\includegraphics[width=0.8\columnwidth]{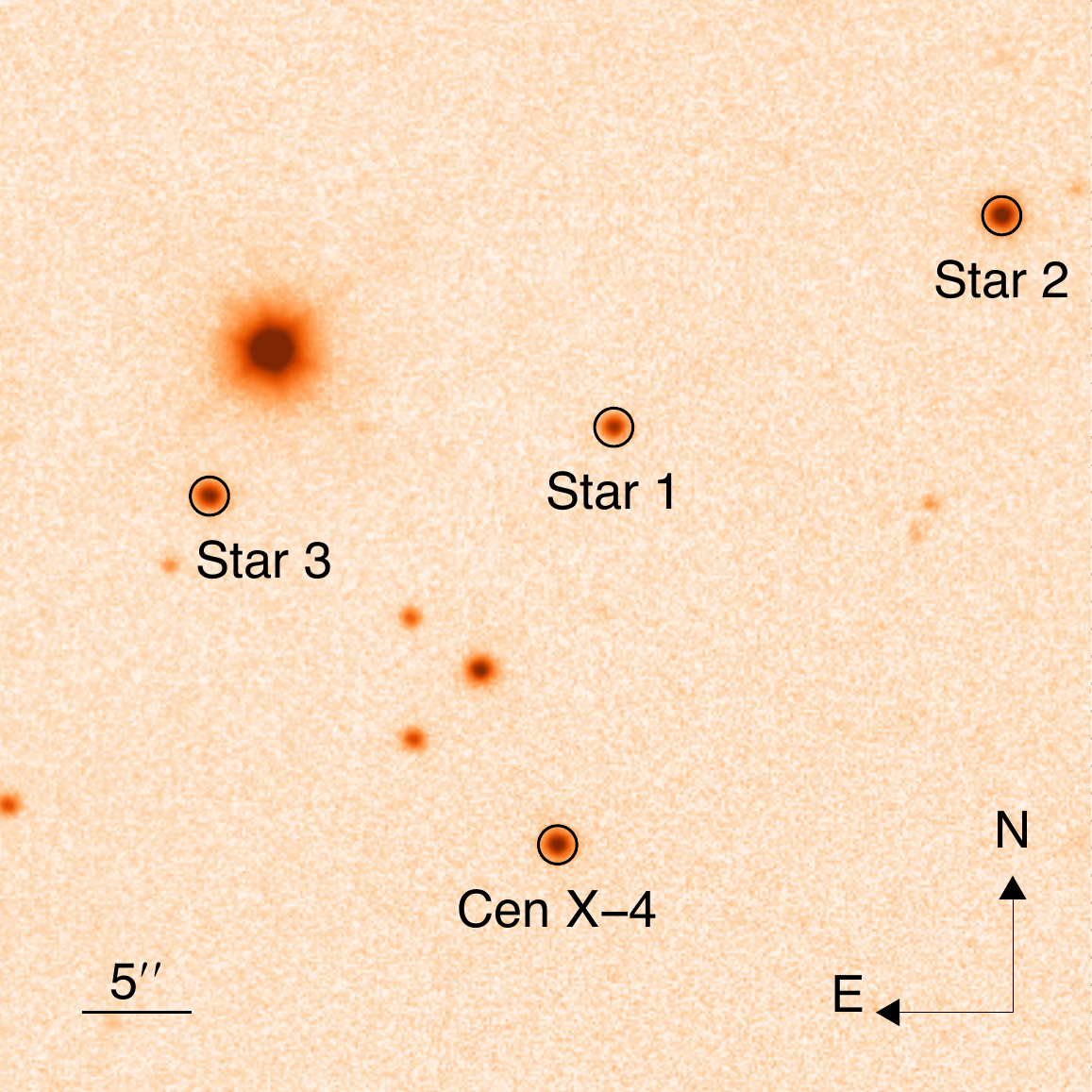}
\caption{PANIC K-band image of Cen~X-4, showing the locations of Cen~X-4 and the three reference stars used.    The image was cropped to show only a $50\arcsec \times 50\arcsec$ region near to Cen~X-4.  The circles have a 0.875\arcsec\ radius.  North is up and East is to the left.}
\label{fig:image}
\end{figure}

To obtain the source flux, we used aperture photometry to acquire source counts and then differential photometry
to acquire relative fluxes for Cen~X-4 and the reference stars. A source radius of 0.875\arcsec\ was used for the aperture photometry.  Table \ref{tab:radec} lists 2MASS positions \citep{2mass} of each source, including Cen~X-4 and the three reference stars used for the differential photometry. We show the locations of the reference stars in Figure~\ref{fig:image}.  This image shows one combined series of dice-5 exposures, cropped to just show the region around Cen~X-4.

\begin{figure}
\centering
\includegraphics[width=0.9\columnwidth]{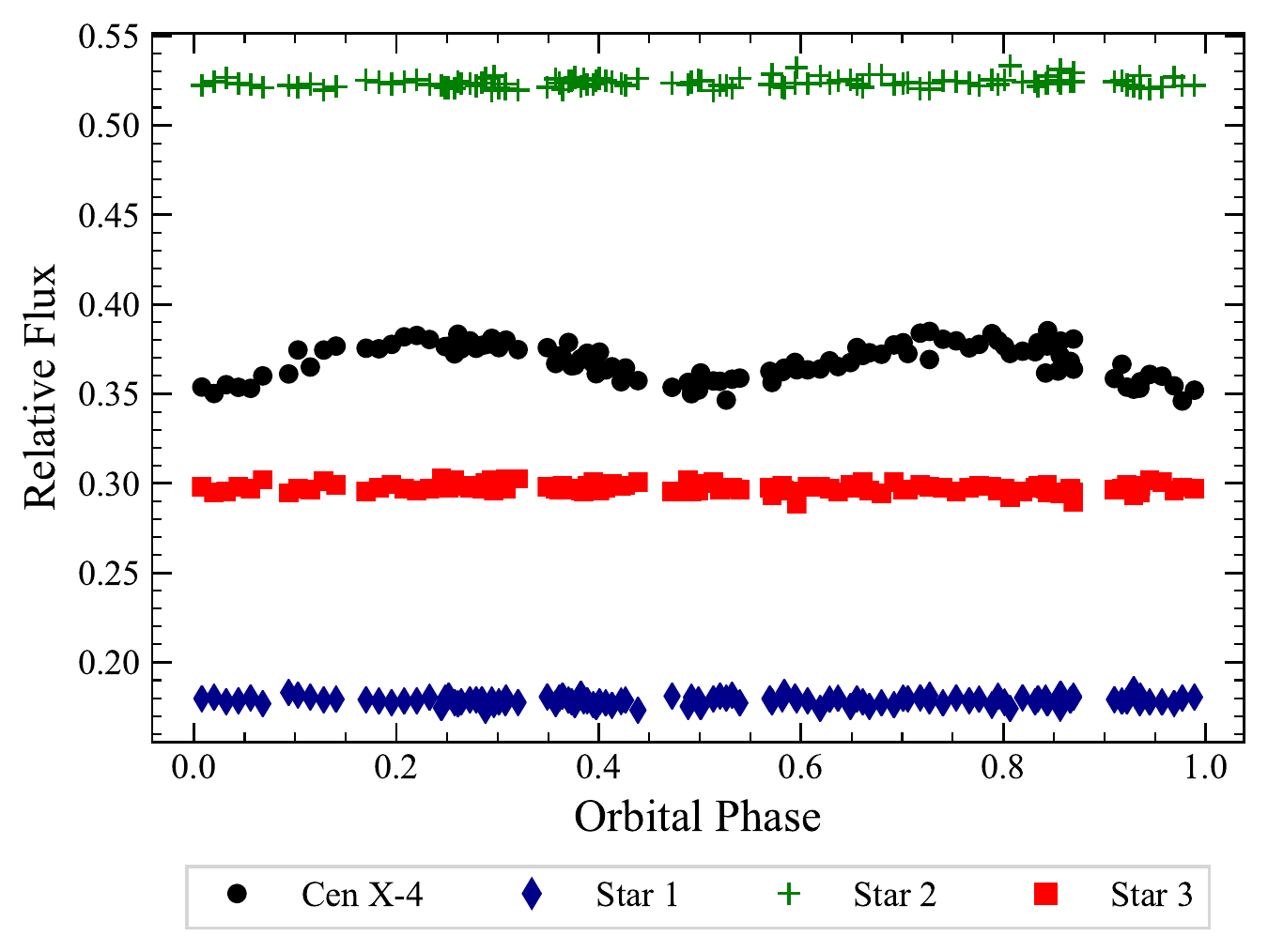}
\caption{Phase-folded K-band lightcurve of Cen X-4 (black circles), along with reference star 1 (blue diamonds), star 2 (green crosses) and star 3 (red squares).  Error bars have been omitted for clarity.}
\label{fig:relphotom}
\end{figure}

The relative fluxes of each source were then converted to apparent magnitudes using 2MASS magnitudes. Photometric uncertainties were calculated using both statistical uncertainties from the aperture photometry and the standard deviation of the reference stars over all images added in quadrature. In order to account for the highest possible uncertainty we chose the reference star with the highest variability, star 1, whose relative standard deviation (RSD) in $K$  is 1.17\%, to calculate uncertainties. Stars 2 and 3 have $K$-band RSDs of 0.49\% and 0.89\%, respectively. Figure~\ref{fig:relphotom} shows the relative K-band flux of the reference stars compared to Cen~X-4.

\begin{figure}	
    \includegraphics[width=0.9\columnwidth]{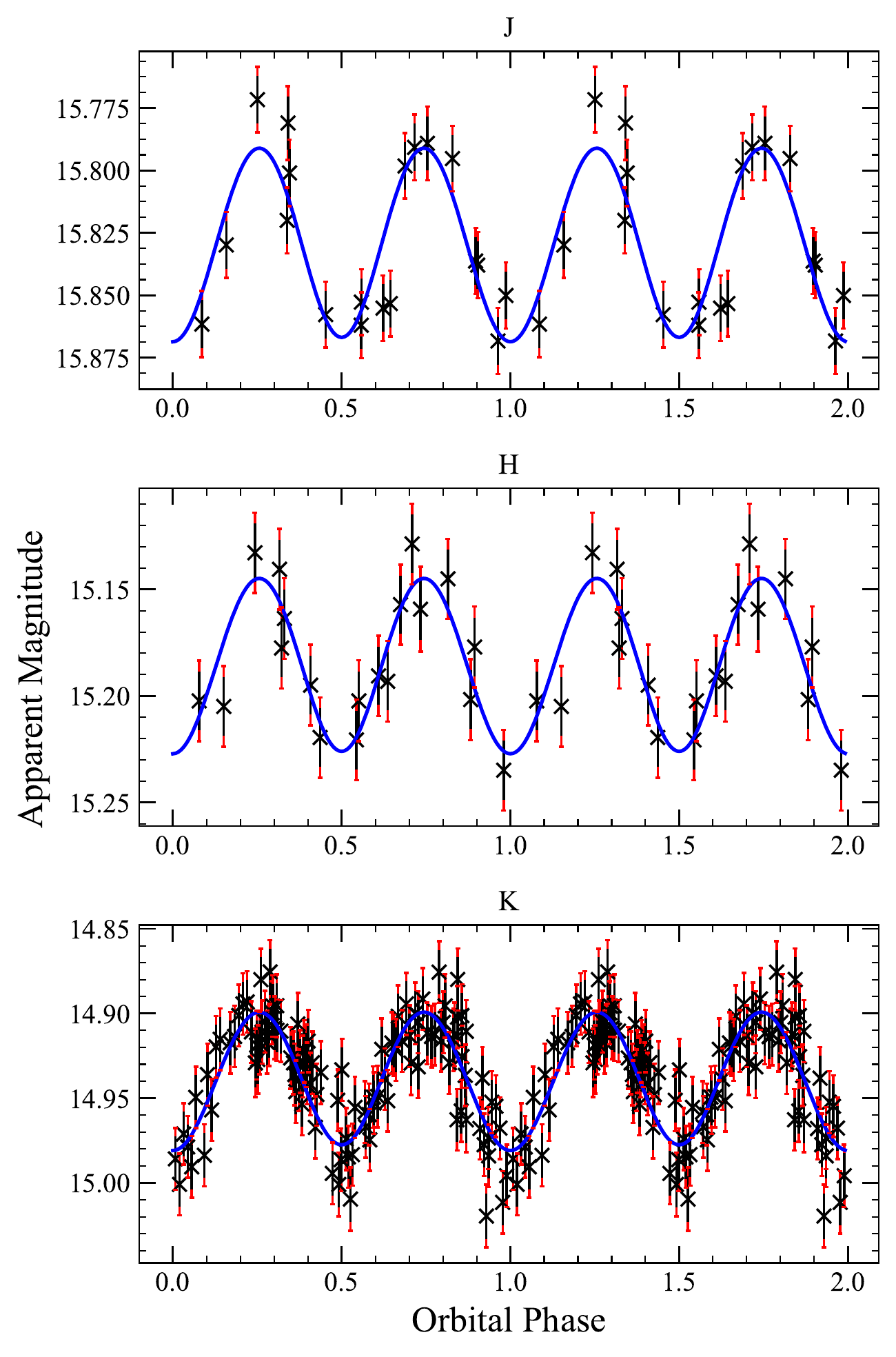}
    \caption{The phase-folded $J$, $H$, and $K$ band light curves. The model is shown here as a blue line and the light curve points have been repeated over a second phase for clarity. The black error bars are the original errors calculated during the photometry process, while the extensions in red are the rescaled error bars.}
    \label{fig:pflc}
\end{figure}

We performed phase folding of the light curves using an orbital period of 0.629059 $\pm$ 0.000017 d and $T_0$ = HJD 2454626.6214 $\pm$ 0.0002 from \citet{shahbaz14}. We present the phase-folded light curves of Cen X-4 in the $J$, $H$, and $K$ bands in Fig. \ref{fig:pflc}.

\begin{table}
	\centering
	\caption{2MASS coordinates for of each of the sources used for differential photometry. The reference stars were chosen for their non-variability and brightness compared to Cen~X-4.}
	\label{tab:radec}
	\begin{tabular}{lcc}
		\hline
		Object & RA & Dec\\
		\hline
		Cen X-4 & 14:58:21.92 & $-$31:40:07.4\\
		Star 1 & 14:58:21.72 & $-$31:39:48.5\\
		Star 2 & 14:58:20.31 &  $-$31:39:38.7\\
		Star 3 & 14:58:23.22 & $-$31:39:51.6\\
		\hline
	\end{tabular}
\end{table}

\section{Light Curve Modeling Results}

We use the Eclipsing Light Curve (ELC) code of \citet{orosz00} in order to simultaneously fit a model
to the $J$, $H$, and $K$ Cen~X-4 light curves and obtain an inclination. ELC has a number of parameters for the binary system that affect the shape of the lightcurve, including the inclination, mass ratio, effective temperature of the secondary star, accretion disk radius and accretion disk temperature. A number of parameters are known from previous studies of Cen~X-4, for instance, initially we fix the values of $P_{\rm orb}$, $K_2$, and $q$ to the values determined by \citet{shahbaz14}.

To start with, we fit the light curves using the geneticELC optimizer, which is based on a genetic algorithm from \citet{Charbonneau95}. We find that the best-fitting models all have reduced-$\chi^2$ values that are formally unacceptable (approximately 1.9). Some quiescent LMXBs have shown significant `flickering' not associated with ellipsoidal variations \citep[e.g.][]{matasanchez17}.  While the Cen~X-4 lightcurves are not significantly different from the ellipsoidal variations, the poor reduced-$\chi^2$ may, in part, be caused by similar additional variability, since the comparison stars have a small dispersion. To account for this, we rescale the error bars (by a factor $\sqrt{1.9}$) so that $\chi^2_{\nu, {\rm min}}$ of the best-fit is approximately 1. While this does not significantly change the best-fitting parameters, it has the effect of increasing the uncertainties on the fit parameters. Both the original and rescaled error bars are shown in~Fig.~\ref{fig:pflc}.  Light curve fitting models quoted here use these increased uncertainties.

To fully explore the parameter space, we utilize the Markov Chain Monte Carlo version of ELC called hammerELC, which is an implementation of the emcee Hammer code of \citet{foreman13}, using 200 walkers over 400 generations.  We fit the lightcurves to determine the inclination but also explore the effects of other system parameters, for instance, allowing the effective temperature to vary in some cases. We try a series of models with and without a significant contribution from a standard geometrically-thin, optically thick accretion disk, and also test the effect of allowing the outer disk radius ($R_{\rm disk}$) and inner disk temperature ($T_{\rm disk}$) to vary.  All other parameters of the model were set to the default of the ELC input files. We set $i$ to vary in the range of 20\degr $<$ $i$ $<$ 50\degr. When $T_{e\mathit{ff}}$ is allowed to be free, we set it to vary in the range of 2000 -- 8000 K. In cases where disk parameters are fixed, the default values for $T_{\rm disk}$ and $R_{\rm disk}$ are 30000 K and 0.75, respectively, where $R_{\rm disk}$ is expressed as a fraction of the neutron star's effective Roche lobe radius. When allowed to vary, $T_{\rm disk}$ is constrained to the range 2000 -- 40000 K and $R_{\rm disk}$ to the range 0.0 -- 1.0.

\begin{figure*}	
    \includegraphics[width=0.9\textwidth]{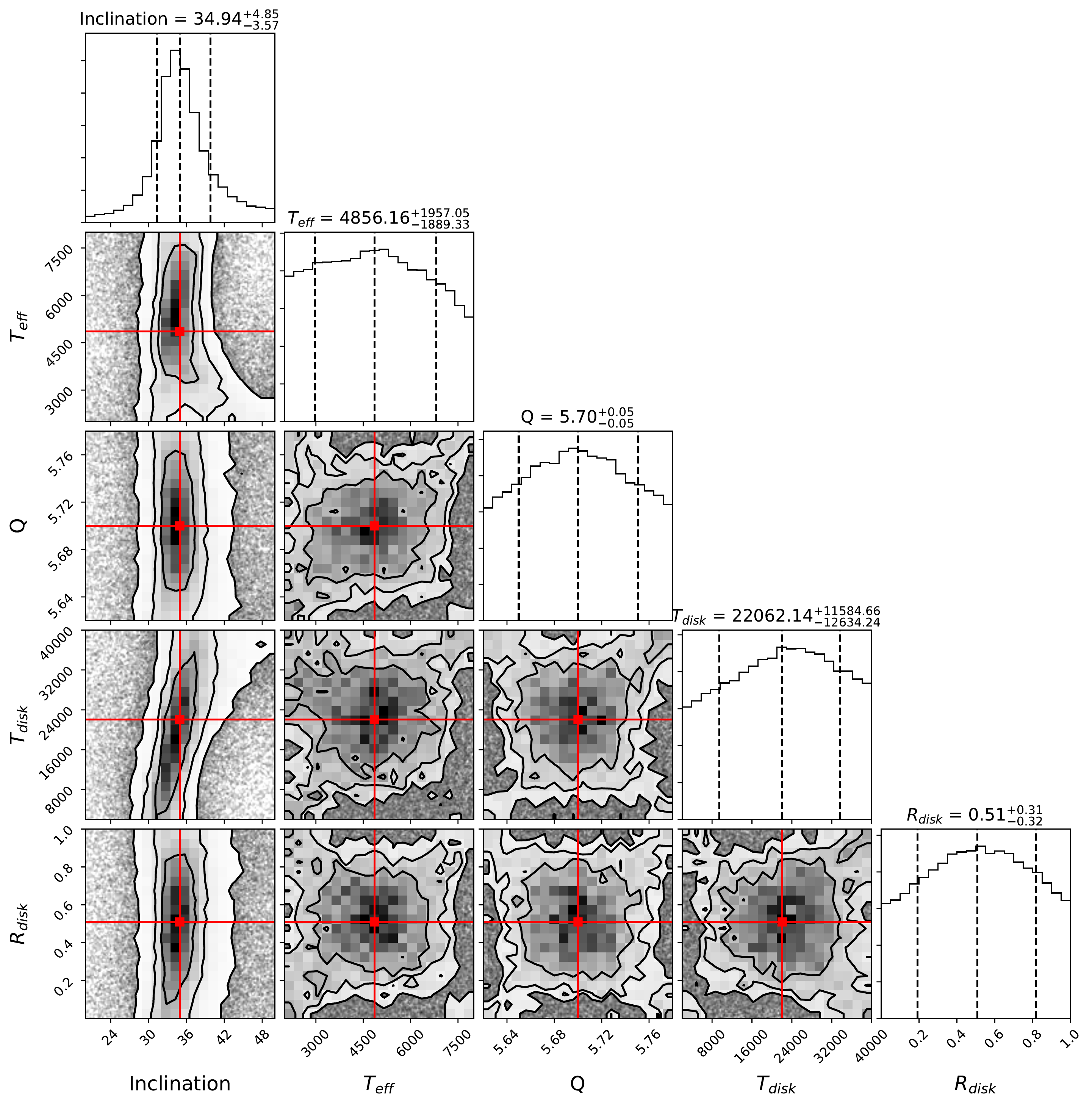}
    \caption{The posterior distributions for model 7 in Table . Contours show the 1, 2 and 3$\sigma$ confidence regions, and red lines show the median of the posterior distributions.  The vertical dashed lines on the histograms show the median of the distribution along with the 1$\sigma$ confidence region.  Note that the mass ratio $Q$ output from ELC is the reciprocal  of $q$ that we use in the text.}
    \label{fig:post}
\end{figure*}

Table \ref{tab:results} shows the parameter values from the various model fits performed. Generally a consistent inclination is found regardless of assumptions made, with the exception of the presence of a disk.  Fits with an accretion disk included systematically lead to higher inclinations around $35^\circ$ (models 3 -- 7).  Model 7 makes the least assumptions about the properties of the binary system. This model varies the inclination, $T_{e\mathit{ff}}$, $q$, $T_{\rm disk}$, and $R_{\rm disk}$, where $q$ is allowed to vary in the range determined by \citet{shahbaz14}. The fit parameters for model 7 are $i = 34.9^{+4.9}_{-3.6}$ degrees, $T_{e\mathit{ff}} = 4860^{+1960}_{-1890}$~K, $q$ = 0.18$\pm$0.1, $T_{\rm disk}=22000^{+12000}_{-13000}$, and $R_{\rm disk}=0.51^{+0.31}_{-0.32}$. The posterior distributions for the parameters of this model can be seen in Fig. \ref{fig:post}. This shows a well constrained inclination.  It also shows a slight degeneracy between the inclination and the disk temperature (lower disk temperatures prefer lower inclinations.  Combining the inclination obtained from this model with the value of the mass function (Equation \ref{eq:fm}) using values of $P_{\rm orb}$, $K_2$, and $q$ calculated by \citet{shahbaz14} gives a neutron star mass of $1.51^{+0.40}_{-0.55}~M_{\sun}$, which is consistent with previous measurements of the mass.

\input{tab_fits}

\section{Discussion}
\subsection{Effects of Model Assumptions}
We modeled the light curves in several different ways in order to better understand the parameter
space and implications of assumptions made during the modeling process. Model 7 makes the least assumptions about the shape and accretion disk contribution to the light curve and gives $i=34.9^{+4.9}_{-3.6}$ degrees, $T_{e\mathit{ff}} = 4900^{+1960}_{-1890}$~K, $q$ = 0.18$\pm$0.1, $T_{\rm disk}=22000^{+12000}_{-13000}$, and $R_{\rm disk}=0.51^{+0.31}_{-0.32}$. This value for $T_{e\mathit{ff}}$ is consistent with $T_{e\mathit{ff}}$ = 4500 K estimated by \citep{gonzalez05}, though, as described by \citet{khargharia10}, there is uncertainty over the donor's spectral type.  \citet{shahbaz14} assume $T_{e\mathit{ff}}$ = 4500 K in their modeling and obtain $32\degr\substack{+8\degr \\ -2\degr}$, consistent with our results.

When we include an accretion disk contribution and allow the effective temperature to be a free parameter while holding other parameters fixed (model 4), we get a value of $T_{e\mathit{ff}} = 4970^{+1770}_{-1730}$~K, also consistent with \citet{gonzalez05}. However, we find that all fits with an accretion disk included systematically lead to higher inclinations. \citet{khargharia10} noted that varying $T_{e\mathit{ff}}$ between 3700 K and 4500 K only changes the inclination by $\sim$1\degr. In our models 1 and 2, and models 3 and 4 we perform fits fixing the effective temperature (models 1 and 3) and allowing it to be free (models 2 and 4), with (models 3 and 4) and without (models 1 and 2) a disk. The difference in inclination from fixing the effective temperature and letting it be optimized in the fit leads to a change of 0.7 and 0.6\degr (without and with a disk, respectively), implying that the effective temperature may not have a large effect on the outcome of the model.  Figure \ref{fig:post} shows that the effective temperature is not well constrained in our modeling results.  The bigger difference comes from whether or not we include an accretion disk, which an increase in inclination of approximately 2 \degr when assuming an accretion disk.

Without taking an accretion disk into account, we find an inclination consistent with previous
estimates obtained through other methods of measuring the inclination. \citet{khargharia10}
obtained near-IR spectroscopy of Cen~X-4 and compared with field K-stars of known spectral type to estimate the fractional donor contribution.  They find the donor star contribution is $0.94\pm0.14$ in the $H$ band and that the accretion disk can contribute up to 10\% of the flux in the $K$ band.  Accounting for this contamination from the disk, \citet{khargharia10}
remodeled the original \citet{shahbaz93} light curves, obtaining $i = 35\degr\substack{+4\degr \\ -1\degr}$, with which our models 3 -- 7 (which also include a disk) agree. \citet{shahbaz14} note that their modeling, which obtained an inclination of $32\degr\substack{+8\degr \\ -2\degr}$,  does not rely on assumptions about contamination from the accretion disk, as long as the presence of a disk doesn't effect the line profile fitting. 
While our best-fitting model agrees with both of these estimates, it is seen from Table
\ref{tab:results} that models that do not account for contamination from the accretion disk have consistently lower inclinations than the models accounting for the disk. It is the systematic uncertainty in the accretion disk contribution that is the limiting factor in determining the inclination and not the statistical uncertainty.  As discussed by \citet{khargharia10}, since the accretion disk contamination could vary over time, contemporaneous near-IR photometry and spectroscopy would be highly beneficial to continuing to improve the inclination measurement.

\subsection{The Accretion Disk in Quiescence}

Since our results are dependent on assumptions about the accretion disk contribution in the near-IR during quiescence, it is worth discussing further. Numerous studies in recent years have observed quiescent variability from the accreting BHs and NSs, ascribed to a dynamically evolving accretion flow \citep{zurita03}. In particular, detailed long term monitoring of the stellar mass black hole A0620$-$00 revealed two distinct modes of quiescent
accretion \citep{cantrell08}, an active mode where the lightcurve was observed to be dominated by a
flickering/flaring component and a passive phase where the lightcurve is dominated by the ellipsoidal
modulations produced by the tidally distorted secondary star.

Observations of the transient neutron star Aql~X-1 have found significant variability (at the 20\%
level) in the $K$-band. This variability was not found to be correlated with the orbital phase of the
system suggesting that the emission originated over a large fraction of the disk \citep{matasanchez17}. The evidence for ongoing accretion related variability in this system is perhaps not so
surprising, as in contrast to Cen~X-4, Aql~X-1 returns to the bright outbursting state every 3 -- 5
years, suggesting a larger mass reservoir that is not drained during each outburst and/or a higher
mass transfer rate from the donor star.

Previous observations have shown evidence for a significant accretion disk contribution to the
continuum flux in the optical in Cen~X-4. \citet{chevalier89} presented optical photometry which detected Cen~X-4
in a clear active state and constrained the disk contribution to be $\sim$ 80/45/25\% in the U/B/V
bands respectively. Recent high time resolution observations show a more modest contribution from
the accretion disk; however, the disk does not appear to contribute significantly in the near-IR bands,
where the disk was measured to contribute only $\sim$6\% of the $I$-band flux \citep{shahbaz10}.
Power-spectrum analysis of these data are consistent with the standard 1/frequency noise spectrum previously
observed from a number of quiescent BHs \citep{zurita03, reynolds07}. Optical spectroscopy has
also revealed many lines consistent with those expected from a disk, i.e.,
H$\alpha,~H\beta,~H\gamma$, He I, He II \citep{shahbaz96}, and a suggested disk contribution of
$\sim$45\% at 4500\AA\ \citep{gonzalez05}.

IR spectroscopy has resulted in the detection of He I and Br$\gamma$ emission lines, consistent with
the presence of ongoing accretion. Further modelling of these spectra suggested that the accretion
disk may contribute $\leq$20\% of the flux in the H-band \citep{khargharia10}. Though, it should be
noted that the accuracy of this estimate is dependent on remaining uncertainty on the nature of the
mass donor star where a subgiant as opposed to a dwarf is consistent with the currently available IR
spectra.

An additional complication is the possible presence of significant spotting on the surface of the
secondary star. \citet{shahbaz14} have detected a significant spot over the polar regions of the
secondary. \citet{cherepashchuk17} have demonstrated how such spotting can introduce variability that
is degenerate with that produced by a variable accretion flow in the BH transient A0620$-$00. As the
donor star in this system is also a late K type star, similar effects may be at play in Cen~X-4.

Unlike the quiescent BHs, we do not expect significant emission from a steady jet in the NS systems
\citep{migliari06}. \citet{baglio14} have presented optical polarimetric observations which do not
detect any significant intrinsic polarization from this system (<0.5\% in the $I$-band). This is
consistent with the non-detection of Cen~X-4 at radio frequencies \citep[$f \leq 14~\mu$Jy,][]{tudor17}.

Finally, longer wavelength observations also place stringent constraints on the contribution of the
disk in the infrared, where both {\it Spitzer} and {\it WISE} have measured fluxes consistent with an
extrapolation of the stellar continuum alone  \citep{muno06, wang14}.  The lack of significant near-IR
disk emission in Cen~X-4 may be a result of the large nature of the last outburst from this system in
1979.

The detection of accretion signatures in previous near-IR spectroscopy suggests that further $J$, $H$, $K$
spectroscopy may be the optimal means to constrain the disk contribution and the precise spectral
type of the secondary star.

\section{Conclusions}
We obtained near-IR photometry of the LMXB Cen~X-4.  Through ellipsoidal modeling of the lightcurves, we constrain the inclination.  The lightcurve fit which makes the least assumptions about the properties of the binary system yields an inclination of $34.9^{+4.9}_{-3.6}$ degrees leading to a neutron star mass estimate of $1.51^{+0.40}_{-0.55}$~$M_{\sun}$. Our values are consistent with previous estimates that do not depend on assumptions about contribution from the disk.

\section*{Acknowledgements}

EKH thanks the National Science Foundation for support through a Research Experience for Undergraduates program at Wayne State University (NSF Grant No. PHY-1460853). We thank Manuel Torres for helpful comments that improved the paper.  We thank the referee, and author of ELC, Jerry Orosz, for a helpful referee report and guidance in using the latest MCMC version of ELC. This paper includes data gathered with the 6.5 meter Magellan Telescopes located at Las Campanas Observatory, Chile.  



\bibliographystyle{mnras}
\bibliography{cenx4}

\bsp	
\label{lastpage}
\end{document}

%% file: tab_fits.tex
\begin{table*}
	
	\caption{Best-fitting parameters from ellipsoidal modeling of the $J$, $H$ and $K$ lightcurves of Cen~X-4.  1-$\sigma$ uncertainties are given.  Each model that we created is listed here along with the free parameters, value of the inclination, and value of any other free parameter included in the model. Here, $i$ is the inclination, $T_{e\mathit{ff}}$ is the effective temperature of the secondary star, $q$ is the mass ratio, and $R_{\rm disk}$ and $T_{\rm disk}$ are the outer radius (as a fraction of the neutron star's effective Roche lobe radius) and temperature of the accretion disk, respectively. $M_{\rm NS}$ is the calculated mass of the neutron star based on the parameters of the model. }
	\label{tab:results}
	\begin{center}
	\begin{tabular}{cccccc}
		\hline
	Model & Free Parameters & Disk Included? & \multicolumn{2}{c}{Parameter Values} & $M_{\rm NS}$ ($M_{\sun}$) \\
	 & & & $i$ (deg) & Other \\
		\hline
		1 & $i$ & No  & $33.1^{+1.0}_{-0.9}$ & -- & $1.78_{-0.14}^{+0.12}$ \\[3pt]
		2 & $i$, $T_{e\mathit{ff}}$ & No  & $33.8^{+2.5}_{-2.4}$ & $5250^{+1650}_{-1950}$ K & $1.68^{+0.31}_{-0.32}$ \\[3pt]
                \hline
		3 & $i$ & Yes  &$35.3\pm1.0$ & -- & 1.49 $\pm$ 0.11 \\[3pt]
		4 & $i$, $T_{e\mathit{ff}}$ & Yes  & $35.9^{+4.3}_{-2.5}$ & $4970^{+1770}_{-1730}$ K & $1.43^{+0.29}_{-0.44}$ \\[3pt]
		5 & $i$, $R_{\rm disk}$ & Yes  & $35.0^{+2.3}_{-2.5}$ & $0.48^{+0.31}_{-0.30}$ & $1.53^{+0.28}_{-0.26}$\\
		6 & $i$, $T_{\rm disk}$ & Yes  & $34.5^{+4.1}_{-3.2}$ & $22000\pm12000$ K & $1.58^{+0.38}_{-0.48}$\\[3pt]
		\hline
		 & $i$ &   & $34.9^{+4.9}_{-3.6}$ &  &  \\[3pt]
		 & $T_{e\mathit{ff}}$ & &  & $4860^{+1960}_{-1890}$ K &  \\[3pt]
		7 & $q$ & Yes  & & $0.18\pm0.01$  &  $1.51^{+0.40}_{-0.55}$ \\[3pt]
		 & $T_{\rm disk}$  & &  & $22000^{+12000}_{-13000}$ K &   \\[3pt]
		 & $R_{\rm disk}$ &  &  & $0.51^{+0.31}_{-0.32}$  &  \\[3pt]
		\hline
	\end{tabular}
	\end{center}
\end{table*}